\documentclass[a4paper,11pt]{article}
\usepackage{pos}

\newcommand{\be}{\begin{equation}}
\newcommand{\ee}{\end{equation}}
\newcommand{\bi}{\begin{itemize}}
\newcommand{\ei}{\end{itemize}}
\title{The de Sitter group and its presence at the late-time boundary}

\author*[a]{Gizem Şengör}

\affiliation[a]{ CEICO, Institute of Physics of the Czech Academy of Sciences,\\
Na Slovance 1999/2, 128 21, Praha 8, Czechia}

\emailAdd{sengor@fzu.cz}

\abstract{Our main goal here is to provide an introduction on some of the well established properties of the representation theory of $SO(d+1,1)$, for those considering to think on physical problems set in de Sitter space in terms of these representations. With this purpose we review two intertwining maps, the map $G$ that is used in constructing a well defined inner product for the complementary series representations and the map $Q$ that is involved in constructing composite representations. We give explicit examples from the late-time boundary of de Sitter on the practical use of the complementary series inner product and in building a tensor product representation from unitary principal series irreducible representations.} 

\FullConference{%
  Corfu Summer Institute 2021 "School and Workshops on Elementary Particle Physics and Gravity"\\
  29 August - 9 October 2021\\
  Corfu, Greece
}

\begin{document}
\maketitle

\section{Introduction}
The group $SO(d+1,1)$ is both the conformal group of Euclidean space in $d$ dimensions and the isometry group of de Sitter in $d+1$ dimensions. It appears in different venues in physics from Euclidean conformal field theory to Cosmology. 

$SO(d+1,1)$ is a noncompact group. It is one of the Lorentz groups along with $ISO(d,1)$ also known as the Poincaré or the inhomogeneous Lorentz group, and $SO(d,2)$. Each one of these groups are involved in physics of maximally symmetric spacetimes with a different value of the cosmological constant. 

Much of the information on the properties of the group $SO(d+1,1)$ and its representations stem from the works of Harish-Chandra. The case of $SO(2,1)$ have been given more special attention. These representations are also captured by the $SL(2,R)$ representations, pedagogical reviews on which can be found in \cite{so21} and \cite{Gelfand}. Recent reviews from the physics literature on the de Sitter group in general dimensions include \cite{Sun} which focus on cases of integer spin, and in sections of \cite{Sengor19} that focus on the scalar representations and how they manifest themselves at the late-time boundary. For the case of spinors we refer the readers to \cite{Pethybridge} and references within. Here we will mainly follow the in depth monologue \cite{Dobrev}. 

In what follows we will focus on the well defined inner product and what it means to have unitary representations. We will explicitly give an example on the use of the scalar complementary series inner product, which involves an intertwining operator $G$, following \cite{Sengor19} with focus on the realization of these representations at the late-time boundary of de Sitter. We will conclude our discussion by pointing out some properties of the tensor product in the case of principal series following \cite{Dobrev} with an example again from the late-time boundary.  

\section{The group $SO(d+1,1)$}
\label{sec:thegroup}
The group $SO(d+1,1)$ is the group of all linear transformations in $d+2$ dimensions that leave the following quadratic form invariant
\be v\eta v=-v_0^2+v_1^2+...+v_d^2+v_{d+1}^2,~~\eta_{\mu\nu}=diag[-1,1,\dots 1].\ee
The group elements $g$, are $(d+2)\times(d+2)$ matrices with unit determinant that also satisfy ${g^0}_0\geq1$, $g^T\eta g=\eta$, where $^T$ stands for the transpose.

The Lie algebra $\mathfrak{so}(d+1,1)$ consists of real $(d+2)\times(d+2)$ dimensional matrices, $X$ that satisfy $X^T\eta+\eta X=0$, with $A,B=0,1,\dots d+1$. With these real generators and real parameters $\alpha$, group elements can be represented by exponentiation of the generators as follows
\begin{align} g_X(\alpha)=e^{\alpha X}\end{align}
As also discussed in \cite{Sun} and \cite{Barut}, with real parameters $\alpha$ and real generators $X$, if this is to be a unitary representation then
\begin{align} g^\dagger_X(\alpha)g_X(\alpha)=1~\implies~X^\dagger=-X.\end{align}
Since $X$ are real their dagger is just the transpose and this implies a basis such that $X_{AB}=-X_{BA}$. In this basis the commutations relations are \cite{Dobrev}
\begin{align}[X_{AB},X_{CD}]=\eta_{AC}X_{BD}+\eta_{BD}X_{AC}-\eta_{AD}X_{BC}-\eta_{BC}X_{AD}.\end{align}
A matrix realization in this basis is 
\begin{align}\left(X_{AB}\right)^C_D=\eta_{AD}\delta^C_B-\eta_{BD}\delta^C_A. \end{align}
These mathematical generators are related to physical generators by $J_{AB}=iX_{AB}$. With the mathematical generators $X$ being real and antisymmetric, the physical generators $J$ are 
\begin{align} \left(J_{AB}\right)^\dagger=\left(iX_{AB}\right)^\dagger=-iX_{BA}=iX_{AB}=J_{AB},\end{align}
guaranteed to be Hermitian.

What the unitarity of the representation implies for the adjoint of the generator has different consequences for each of the cases of $so(d+1,1)$ algebra, of rotation algebra as pointed out in \cite{Barut} and of  $so(d,2)$ algebra as pointed out in \cite{Sun}.

Certain generators among the $\frac{1}{2}(d+1)(d+2)$ generators generate specific subgroups. Here we list them with emphasis on their compact or non-compact nature. 

The non-compact subgroups are :
\begin{itemize}
	\item{$A=SO(1,1)$: The generator $D=X_{d+10}$ generates dilatations.}
	\item{$N$: The generators $C_i=X_{i0}-X_{id+1}$ where $i=1,...,d$, generate special conformal transformations.}
	\item{$\tilde{N}$: The generators $T_i=X_{i0}+X_{id+1}$ generate spatial translations.}
	\item{$H$, all together the generators $D, X_{12},...,X_{\left(2\left[\frac{d}{2}\right]-1\right) \left(2\left[\frac{d}{s}\right]\right)}$ where $\left[\frac{d}{2}\right]$ stands for the integer part of $\frac{d}{2}$ generate the so called Cartan subgroup. This is the abelian group of $\left[\frac{d}{2}\right]+1$ dimensional diagonalizable matrices.}
\end{itemize}
The above are all abelian subgroups. The compact subgroups are:
\begin{itemize}
	\item{$K=SO(d+1)$: This is the maximally compact subgroup. It is generated by $X_{ab}$, where $a,b=1,...,d+1$.}
	\item{$M=SO(d)$: The generators $X_{ij}$ with $i,j=1,\dots d$ generate the so called "Euclidean Lorentz group". This is the subgroup of spatial rotations. This subgroup is the centralizer of $A$ in $K$ this means for $m\in M$ and $a\in A$, $mam^{-1}=a$ where $m^{-1}$ denotes the inverse of $m$.}
	\end{itemize}
The quadratic Casimir operator of this algebra is
\begin{align}C_2(\chi)=-\frac{1}{2}X^2_{ij}+D^2+dD+C_iT_i,\end{align}
with eigenvalues
\begin{align}
c_2(\chi)=l(l+d-2)+c^2-\frac{d^2}{4}.
\end{align}
We explain the labels $l$ and $c$ in the next section.
 
\section{The unitary irreducible representations of $SO(d+1,1)$}
\label{sec:UIR}
Representations of a group are labelled by the eigenvalues of the quadratic casimir of the algebra. For $\mathfrak{so}(d+1,1)$, the free variables of these eigenvalues are spin, here denoted by $l$, which labels the representations of the rotation subgroup $M$, and the scaling weight, denoted by $c$, which is associated to the representations of dilatations and special conformal transformations. 

Primary operators transform under dilatations with a scaling dimension $\Delta$ as follows
\be \mathcal{O}(\lambda x)=\lambda^{-\Delta}\mathcal{O}(x).\ee
For the group $SO(d+1,1)$, the scaling dimension has the following form
\be \Delta=\frac{d}{2}+c.\ee
A part of the scaling dimension is fixed by the number of spatial dimensions\footnote{The quantity $\frac{d}{2}$ corresponds to the half sum of the restricted positive roots \cite{Dobrev}.}. The free part $c$, is called the \emph{scaling weight} and it can be either real or purely imaginary. The unitary irreducible representations fall under four different categories, characterized by the scaling weight and spin. These categories are denoted by $\chi=\{l,c\}$. 

A unitary representation preserves the inner products of states. There are two ways to build these states. One way is to construct the states $|l,c\rangle$, by considering how the generators of the algebra act on them. A second way is to first build operators $\mathcal{O}_{\{l,c\}}$ from functions on finite group elements. Then states are defined by considering the action of these operators on the vacuum $|0\rangle$ that remains invariant under $SO(d+1,1)$ transformations, $|\mathcal{O}_{\{l,c\}}\rangle\equiv\mathcal{O}_{\{l,c\}}|0\rangle$. In either way, the normalization for the states works differently for each category.

Following the first route of construction from the algebra, when one constructs normalized states 
\be |l,c\rangle_N\equiv \mathcal{N}|l,c\rangle,\ee 
such that
\be  _N\langle l',c' |l,c\rangle_N=\delta_{l,l'}\delta_{c,c'} \ee
the normalization $\mathcal{N}$ has a different value for each category. In \cite{Barut} this is worked out explicitly for $SO(2,1)$ in comparison with $SO(3)$. From more recent literature, \cite{chargedads2} and \cite{invitation} also pursue this route, with focus on the principal series category, for $SL(2,R)$ representations which accommodate $SO(2,1)$ representations.

In the second route of construction from function spaces the well defined inner product for each category differs. Here following \cite{Dobrev} we will summarize this construction with focus on scalar representations. But before moving on to details let us first briefly describe each category. 

A major distinction between categories is to do with the scaling weight being purely imaginary or real. In the case of purely imaginary scaling weight, ($c=i\rho$, with $\rho\in \mathbb{R}$), the inner product is straight forward and this hosts only the \emph{principal series representations}. In the case where the scaling weight is real, there are three different categories, namely \emph{complementary series}, \emph{exceptional series} and \emph{discrete series}. For these three categories the inner product involves intertwining operators. Each of these three categories span a different range of the real scaling weight and each one has a different intertwining operator involved. The range on $c$ and the intertwining operator involved depends on spin. 

\bi
\item{{\bf Principal Series:}} This is the category with purely imaginary scaling weight and straight forward inner product as we will explain shortly.
\ei

\bi
\item{{\bf Complementary Series:} For $l=0$ this category arises in the range $-\frac{d}{2}<c<\frac{d}{2}$ provided $\frac{d}{2}\geq1$, and for $l=1,2,..$, in the range $1-\frac{d}{2}<c<\frac{d}{2}-1$ with $\frac{d}{2}>1$. The well defined inner product involves an intertwining operator, $G_\chi$. This is a normalizable operator that maps representations $\chi=\{l,c\}$ to their equivalent duals $\tilde{\chi}=\{l,\tilde{c}=-c\}$, leaving the character invariant.}
\ei

\bi
\item{{\bf Exceptional Series:} These representations appear at the values of $c$ for which the complementary series range stops. At these points the complementary series intertwining operator becomes ill defined, as we will discuss in section \ref{subsec:complementary series inner product}. For instance for $l=0$, $c=-\frac{d}{2}$ belong to exceptional series in any dimensions. In general for exceptional series, the dual representations $\chi$ and $\tilde{\chi}$ are not equivalent. This category further splits into four different categories and they are reducible representations. They involve different intertwining operators. These intertwining operators have explicit forms in momentum space for some of the categories and explicit forms in position space for others.}
\ei

\bi
\item{{\bf Discrete Series:}} These representations arise when the rank of the group equals the rank of the maximally compact subgroup ($rank SO(d+1,1)=rank SO(d+1)$), this is only satisfied for $d+1=even$. Moreover, unitary irreducible discrete series representations are not unitarily equivalent to their mirror images\footnote{
	Mirror images of representations $D^l(m)$ of the rotation subgroup $M$ are defined as\cite{Dobrev}
	\begin{align}D^{\Pi l}(m)\equiv D^{l}(\Pi m \Pi^{-1})\end{align}
where $\Pi$ is the space reflection with properties $\Pi^2=\Pi^T\Pi=1$, $det(\Pi)=-1$. }, and this further restricts them to exist only for $d+1=2,4$ dimensions \cite{Dobrev}. 
\ei
In the case of two dimensions, $SO(2,1)$ representations are accommodated within the $SL(2,R)$ representations. In two dimensions, $M=SO(1)$, the rotations are trivial and the representations are labelled only by $c$. The $SL(2,R)$ representations are constructed on spaces of homogeneous functions of real variables with degree $s$ which can have even or  odd parity $\epsilon$ (where $\epsilon=0$ for even, $\epsilon=1$ for odd parity). These representations are labelled by $\{\epsilon,s\}$ \cite{Gelfand}. The $\{\epsilon=0,s=-2c\}$ representations of $SL(2,R)$ correspond to the elementary $SO(2,1)$ representations (see for instance \cite{Dobrev} appendix B.4).

Elementary representations are induced by the stability subgroup of the group of interest. In the case of $SO(d+1,1)$ the stability subgroup is the \emph{parabolic subgroup} which is the combination of special conformal transformations, dilatation and rotations, $P=NAM$. For comparison, the parabolic subgroup for $SO(d+1,1)$ plays the role of the \emph{Little group} for Poincaré. In building representations based on functions $\mathfrak{f}(g)$, from finite group elements $g\in G$ to a Hilbert space, the representation $\mathcal{T}^\chi$ acts on these functions by a homomorphism. The functions $\mathfrak{f}(g)$ form a function space $\mathcal{C}_{\chi}$, with certain properties. In the case of $SO(d+1,1)$, these functions have their values on the Hilbert space $\mathcal{V}^{l}$ on which the unitary irreducible representations $(D^l(m))$, of the rotation subgroup $M=SO(d)$ are realized. The elements of $\mathcal{C}_\chi$ are infinitely differentiable and satisfy a so called \emph{covariance condition}. The covariance condition is an identity that determines how the function scales if one considers the specific argument $gnam$, that is the combination of an arbitrary group element $g$ with elements from the subgroups of special conformal transformations $(n\in N)$, dilatation $(a\in A)$ and rotations $(m\in M)$:
\begin{align}\label{covcond}
\text{irrep covariance condition:}~ \mathfrak{f}(gnam)=|a|^{\frac{d}{2}+c}D^l(m)^{-1}\mathfrak{f}(g).\end{align}

There is a one-to-one correspondance between functions of group elements $\mathfrak{f}(g)$ and functions on position space $f(x)$ where $x\in \mathbb{R}^d$ as follows: \cite{Dobrev} 

\begin{subequations}
		\label{grouptoposition_irrep}
	\begin{align}
	\text{to each}~x\in \mathbb{R}^d~\text{corresponds a unique}~\tilde{n}_x\in \tilde{N}~\text{such that:}&~f(x)=\mathfrak{f}(\tilde{n}_x)\\
	\text{to each}~g\in G~\text{corresponds a unique}~x_g\in \mathbb{R}^d~\text{such that:}&~g^{-1}\tilde{n}_{x}=\tilde{n}_{x_g}n^{-1}a^{-1}m^{-1}
	\end{align}
\end{subequations}

where the subscripts are to emphasize the uniquely corresponding element. For instance $\tilde{n}_{x}$ is the element of translations corresponding to a specific $x$,  and the second line is the defining condition for $x_g$, the element of $\mathbb{R}^d$ corresponding to a specified $g$.  
Due to dilatations and special conformal transformations the volume elements are related as
\begin{align}d^dx_g=|a|^{-d}d^dx.\end{align}
To summarize so far, we have the function space
\begin{align}\mathcal{C}\chi=\Bigg\{\mathfrak{f}:G\to \mathcal{V}^l~\text{such that}~\mathfrak{f}(gnam)=|a|^{\frac{d}{2}+c}D^l(m)^{-1}\mathfrak{f}(g)\Bigg\}.\end{align}
Representations of $SO(d+1,1)$ act on functions  that belong to this function space by the following homomorphism
\begin{align}\label{rep}[\mathcal{T}^\chi(g)\mathfrak{f}](g')=\mathfrak{f}(g^{-1}g')~\text{where}~g,g'\in G,~\mathfrak{f}\in \mathcal{C}_\chi.\end{align}
One can further complete the function space $\mathcal{C}_\chi$ to a Hilbert space by equipping it with an inner product. This inner product can be build upon the inner product $\langle.|.\rangle$ that is invariant under rotations $M$, and can be expressed in position space as follows
\begin{align}
\label{princinnerproduct}\left(f_1,f_2\right)=\int d^d x_g\langle f_1(x_g)|f_2(x_g)\rangle.
\end{align}
If $\mathcal{T}^\chi(g)$ is a unitary representation, then it should preserve the inner product
\begin{align}
\label{unitarity}\text{unitarity:}~\left(\mathcal{T}^\chi(g)f_1,\mathcal{T}^\chi(g)f_2\right)=\left(f_1,f_2\right).
\end{align}
By considering $g'=\tilde{n}$ in \eqref{rep} and \eqref{covcond} one can re-express the homomorphism and the covariance condition for functions on position space. Through the properties of $\mathcal{C}_\chi$ we mentioned so far the left hand side of \eqref{unitarity} gives \cite{Sengor19}  
\begin{align} \left(\mathcal{T}^\chi(g)f_1,\mathcal{T}^\chi(g)f_2\right)=\int d^dx_g|a|^{-(c^*+c)}\left(f_1(x_g),f_2(x_g)\right)\end{align}
where $^*$ denotes complex conjugation. Only in the case of principal series representations the contribution $|a|^{-(c^*+c)}$ disappears due to $c$ being purely imaginary. The inner product \eqref{princinnerproduct}, works for principal series representations. For the other three representations intertwining operators, which flip the sign of the scaling weight while leaving $l$ invariant, are introduced so as to remove this piece. Now let us see how this works with an explicit example for the case of complementary series representations.

\subsection{The complementary series inner product}
\label{subsec:complementary series inner product}
The well defined inner product for the complementary series is
\begin{align}
\left(f_1,G_{\tilde{\chi}}f_2\right)
\end{align}
where $G_{\tilde{\chi}}$ is an intertwining operator. The intertwining operator is an invertible map between the function space $\mathcal{C}_\chi$ and $\mathcal{C}_{\tilde{\chi}}$ where $\chi=\{l,c\}$ while $\tilde{\chi}=\{l,\tilde{c}\}$ such that $\tilde{c}=-c$. For $G_{\chi}$, its inverse is $G_{\tilde{\chi}}$. It maps representations with $\Delta=\frac{d}{2}+c$ to representations with $\tilde{\Delta}=\frac{d}{2}-c$. The scaling weight enters into the eigenvalue of the quadratic casimir by $c^2$ and therefore the intertwined operators belong to the same Casimir eigenvalue. Moreover the intertwining operation is a similarity operation. Hence the intertwined representations have the same trace and are equivalent representations. Since under this operation the dimensions satisfy
\begin{align} \Delta+\tilde{\Delta}=d\end{align}
the intertwining operator carries out a shadow transformation.

 In employing these intertwining operators one needs to pay attention to the normalization and on which function space which operator acts. $G_\chi$ acts on $\mathcal{C}_{\tilde{\chi}}$ and is well defined for $Re(c)<0$, while $G_{\tilde{\chi}}$ acts on $\mathcal{C}_{\chi}$ and is well defined for $Re(c)>0$. For the normalization of the intertwining operator, there are a few possible choices each of which is suitable for a different purpose. The appropriate normalization for the positivity of the scalar product is
\begin{align}\label{nplus}n_+(\chi)=n_+(l,c)=\left(\frac{d}{2}+l+c-1\right)\frac{\Gamma\left(\frac{d}{2}+c-1\right)}{\Gamma(-c)}.\end{align}
Note that the normalization choice $n_+(\chi)$ diverges for certain values of $c$, for instance whenever the gamma function has a negative integer argument. This happens for the exceptional series representations and the intertwining operator has a different normalization in those cases. The appropriate normalization for Wightman positivity and operator product expansion is also different then \eqref{nplus}. These other normalizations can be found in \cite{Dobrev} section 5.C.

One place where one encounters the $SO(d+1,1)$ representations is in considering the late-time behaviour of scalar fields in the Poincaré patch of de Sitter \cite{Sengor19}, with metric
\begin{align}\label{metric} ds^2=\frac{-d\eta^2+d\vec{x}^2}{H^2|\eta|^2},~~\eta\in(-\infty,0),~\vec{x}\in\mathbb{R}^d.
\end{align}
 In the late-time limit free quantized scalar fields that satisfy Bunch-Davies initial conditions with masses in the range $m<\frac{d}{2}H$, decomposed in terms of Fourier modes behave as 
\begin{align}
\lim_{\eta\to 0}\phi(\vec{x},\eta)&=\int \frac{d^2k}{(2\pi)^d}\left[|\eta|^{\frac{d}{2}-\nu}\alpha^L(\vec{k})+|\eta|^{\frac{d}{2}+\nu}\beta^L(\vec{k})\right]e^{i\vec{k}\cdot\vec{x}}
\end{align}
with
\begin{align}
\nu^2&=\frac{d^2}{4}-\frac{m^2}{H^2},\\
\alpha^L(\vec{k})&=-\frac{i}{\pi}\Gamma(\nu)N_\alpha \left[a_{\vec{k}}-a^\dagger_{-\vec{k}}\right]\left(\frac{k}{2}\right)^{-\nu}\\
\beta^L(\vec{k})&=\frac{N_\beta}{\Gamma(\nu+1)}\left[\left(1+icot(\pi\nu)\right)a_{\vec{k}}+\left(1-icot(\pi\nu)\right)a^\dagger_{-\vec{k}}\right]\left(\frac{k}{2}\right)^{\nu}.
\end{align}
 In what follows, our notation is such that $\nu$ is positive. The solutions with Bunch-Davies initial conditions are Hankel functions and the above expressions arise from the asymptotic behaviour of Hankel functions \cite{Sengor19, Sengor21}. Here $a_{\vec{k}}$ and $a^\dagger_{\vec{k}}$ are annihilation and creation operators. They satisfy the following commutation rule
\begin{align}
\left[a_{\vec{k}},a^\dagger_{\vec{k}'}\right]=(2\pi)^d\delta^{(d)}\left(\vec{k}-\vec{k}'\right),
\end{align}
and act on the vacuum $|0\rangle$ as follows
\begin{align} &a_{\vec{k}}|0\rangle=0,~~a^\dagger_{\vec{k}}|0\rangle=|\vec{k}\rangle,\\
&\langle\vec{k}|\vec{k}'\rangle=(2\pi)^d\delta^{(d)}\left(\vec{k}-\vec{k}'\right).\end{align}
As such  $\alpha^L(\vec{k})$ and $\beta^L(\vec{k})$ are operators at the late-time boundary of de Sitter which have nontrivial commutation relations. $N_\alpha$ and $N_\beta$ are the normalizations which we will now discuss. By checking what happens to these solutions under dilatations one can identify that the scaling weights for $\alpha^L(\vec{k})$ and $\beta^L(\vec{k})$ are respectively $c_\alpha=-\nu$ and $c_\beta=\nu$ \cite{Sengor19}. For the range $0<m<\frac{d}{2}H$ the late-time operators correspond to complementary series representations.

With $n_+$ normalization the intertwining operator and its inverse for $l=0$ in momentum space are \cite{Dobrev}, \cite{Sengor19}
\begin{align}
\text{for $Re(c)<0$:}&~~G_\chi:\mathcal{C}_{\tilde{\chi}}\to\mathcal{C}_{\chi}~~\text{with}~~G^+_{\{0,c\}}(k)=\left(\frac{k^2}{2}\right)^c,\\
\text{for $Re(c)>0$:}&~~G_{\tilde{\chi}}:\mathcal{C}_{\chi}\to\mathcal{C}_{\tilde{\chi}}~~\text{with}~~G^+_{\{0,\tilde{c}\}}(k)=\left(\frac{k^2}{2}\right)^{\tilde{c}}=\left(\frac{k^2}{2}\right)^{-c}.
\end{align}
For $\beta^L(\vec{k})$ since $c_\beta=\nu>0$ its shadow dual is obtained from
\begin{align}\tilde{\beta}^L(\vec{k})=G_{\tilde{\chi}}\beta^L(\vec{k}),\end{align}
while for $\alpha^L(\vec{k})$ with $c_\alpha=-\nu<0$
\begin{align}\alpha^L(\vec{k})=G_{\chi}\tilde{\alpha}^L(\vec{k}).\end{align}

Defining $\Omega=\int \frac{d^dk}{(2\pi)^d}\langle\vec{k}|-\vec{k}\rangle$ and demanding these operators be normalized up to a dirac delta function, such that
\begin{align}\frac{1}{\Omega}\left(\mathcal{O},\tilde{\mathcal{O}}\right)=\frac{1}{\Omega}\int\frac{d^dk}{(2\pi)^d}\langle 0|\mathcal{O}(\vec{k})\tilde{\mathcal{O}}(\vec{k})|0\rangle\overset{!}{=}1\end{align}
we obtain the following normalized late-time operators
\begin{align}
\alpha^L_N(\vec{k})&=-i2^{\nu/2}\left[a_{\vec{k}}-a^\dagger_{-\vec{k}}\right]k^{-\nu}\\
\beta^L_N(\vec{k})&=2^{-\nu/2}\left[\frac{1+i\cot(\pi\nu)}{1-i\cot(\pi\nu)}a_{\vec{k}}+a^\dagger_{-\vec{k}}\right]k^{\nu},
\end{align}
 and their shadows \cite{Sengor21}
\begin{align}
\tilde{\alpha}^L_N(\vec{k})&=-i2^{-\nu/2}k^\nu\left[a_{\vec{k}}-a^\dagger_{-\vec{k}}\right],\\
\tilde{\beta}^L_N(\vec{k})&=2^{\nu/2}k^{-\nu}\left[\frac{1+i\cot(\pi\nu)}{1-i\cot(\pi\nu)}a_{\vec{k}}+a^\dagger_{-\vec{k}}\right].
\end{align}
The nontrivial commutation relations we mentioned above in passing are 
\begin{align}
\left[\beta^L_N(\vec{k}),\alpha^L_N(\vec{k}')\right]=\frac{2i}{1-i\cot(\nu\pi)}(2\pi)^d\delta^{(d)}(\vec{k}+\vec{k}')=\left[\tilde{\beta}^L_N(\vec{k}),\tilde{\alpha}^L_N(\vec{k}')\right].
\end{align}
At the level of states obtained from these operators by
\begin{align} |\mathcal{O}(\vec{k})\rangle\equiv\mathcal{O}(\vec{k})|0\rangle,\end{align}
operators $\alpha^L_N(\vec{k})$ and $\tilde{\beta}^L_N(\vec{k})$ lead to the same state
\begin{align}|\alpha^L_N(\vec{k})\rangle=i2^{\nu/2}k^{-\nu}|-\vec{k}\rangle=i|\tilde{\beta}^L_N(\vec{k})\rangle, \end{align}
but at the level of operators they are not equal to each other and they do not commute either
\begin{align}\left[\tilde{\beta}^L_N(\vec{k}),\alpha^L_N(\vec{k}')\right]=\frac{2i}{1-i\cot(\nu\pi)}2^\nu k^{-2\nu}(2\pi)^d\delta^{(d)}(\vec{k}+\vec{k}'). \end{align}Similar argument also holds for $\tilde{\alpha}^L_N(\vec{k})$ and $\beta^L_N(\vec{k})$.

Another example is the work of \cite{Mourad} where they construct field operators in position space for the complementary series representations by incorporating the properties of the complementary series inner product in the definition of position space annihilation and creation operators.

\section{Composite states}
\label{sec:composite}
In the previous section we were interested in irreducible representations. The irreducible representations, as we saw in section \ref{sec:UIR},  are induced by the subgroup $P=NAM$, the stability subgroup of $G=SO(d+1,1)$. Irreducible representations involve functions that act on a single point. The composite representations on the other hand, involve two noncoinciding points. The stability subgroup of $SO(d+1,1)$ acting on two noncoinciding points $(x_1,x_2)$, such that $x_1\neq x_2$, on the Euclidean space is isomorphic to the subgroup $MA$ (Lemma 9.1 in \cite{Dobrev}). Thus the composite representations are induced by the subgroup $MA$. We would like to end our discussion with a brief review of how composite reducible representations can be obtained from the irreducible ones. This section mostly points out key features from  chapter 9 of \cite{Dobrev}. 

There are two complementary concepts one would like to understand of composite representations. One question is how to put together two irreducible representations into a representation. This is done via the Kronecker product, also known as tensor product, of irreducible representations. The second question is how to reduce a given representation into its irreducible components. In essence the tensor product involves functions of two entries from $G\times G$ that have values in the product space $\mathcal{V}^{l_1}\otimes \mathcal{V}^{l_2}$. The question of reduction on the other hand involves being able to write functions that have a single argument from $G$ and values in a product space, where the main endeavour is in understanding what this product space can be. We will discuss these two questions in their separate sections. 

\subsection{The tensor product of two irreducible representations }

Given two irreducible representations $\chi_1=\{l_1,c_1\}$ and $\chi_2=\{l_2,c_2\}$ that act on function spaces $\mathcal{C}_{\chi_1}$ and $\mathcal{C}_{\chi_2}$, the tensor product representation acts on the function space $ \mathcal{C}_1\otimes\mathcal{C}_2$, let us denote this product function space by $\mathcal{C}_{\chi_1\otimes\chi_2}$. This is a space of infinitely differentiable functions $\mathfrak{f}(g_1,g_2)$ from $G\times G$ to $\mathcal{V}^{l_1}\otimes \mathcal{V}^{l_2}$ that satisfy the following covariance condition
\begin{align}
\nonumber&\text{tensor product covariance condition:}\\ \label{covcond_product}&~~~\mathfrak{f}(g_1p_1,g_2p_2)=\left[D^{\chi_1}(p_1^{-1})\otimes D^{\chi_2}(p_2^{-1})\right]\mathfrak{f}(g_1,g_2)~\text{for}~p_1,p_2\in MAN,~g_1,g_2\in G.
\end{align} 
Here $D^{\chi}(p)$ is short hand for $D^{\chi}(ma)=|a|^{-\frac{d}{2}-c}D^l(m)$ that one can recognize in section \ref{sec:UIR}. The composite representation acts on these functions by the following homomorphism
\begin{align}
\left[\mathcal{T}^{\chi_1\otimes\chi_2}(g)\mathfrak{f}\right](g_1,g_2)=\mathfrak{f}(g^{-1}g_1,g^{-1}g_2)~\text{with}~g,g_1,g_2\in G, \mathfrak{f}\in \mathcal{C}_{\chi_1\otimes\chi_2}.
\end{align}
There is again a unique correspondence between functions on group elements $\mathfrak{f}(g_1,g_2)$ and functions on position space $f(x_1,x_2)$ estabilished by the following identities \cite{Dobrev}
\begin{subequations}
		\label{grouptoposition_compositerep}
	\begin{align}
\nonumber	\text{to each}~(x_1,x_2)\in \mathbb{R}^d\times \mathbb{R}^d\text{such that}~x_1\neq x_2&,~\text{corresponds a unique}~(\tilde{n}_{x_1},\tilde{n}_{x_2})\in \tilde{N}\times\tilde{N}~\text{such that:}\\
f(x_1,x_2)&=\mathfrak{f}(\tilde{n}_{x_1},\tilde{n}_{x_2})\\
\nonumber	\text{to each}~g\in G~\text{corresponds a unique}~~ p(x,g)&\in MAN~\text{defined by:}\\
g^{-1}\tilde{n}_{x}&=\tilde{n}_{g^{-1}x}p(x,g)^{-1}
	\end{align}
\end{subequations}
which are similar to those in \eqref{grouptoposition_irrep}, involved in the case of irreducible representations.

In the case of both $\chi_1$ and $\chi_2$ being in the principal series the composite representation is unitary with respect to the following inner product \cite{Dobrev}
 \begin{align}\label{compositeinner}
 \left(f_1,f_1\right)=\int dx_1 dx_2 \langle f_1(x_1,x_2),f_2(x_1,x_2)\rangle,
\end{align}
where this time $\langle .,.\rangle$ is the $M-$invariant inner product on $\mathcal{V}^{l_1}\otimes\mathcal{V}^{l_2}$.

A case where principal series representations of $SO(d+1,1)$ appear is in the late-time behaviour of free scalar fields on de Sitter that satisfy Bunch Davies initial conditions with masses in the range $m>\frac{d}{2}H$. One late-time operator among the principal series representations, normalized with respect to the principal series inner product  is \cite{Sengor21}
\begin{align}
\alpha^H_N(\vec{k})&=\sqrt{\rho\pi sinh(\rho\pi)}\left[-i\frac{\Gamma(i\rho)}{\pi}e^{-\rho\pi}a_{\vec{k}}+\frac{1}{sinh(\rho\pi)\Gamma(1-i\rho)}a^\dagger_{-\vec{k}}\right]\left(\frac{k}{2}\right)^{-i\rho},\\
\text{where}~\rho^2&=\frac{m^2}{H^2}-\frac{d^2}{4}.
\end{align}
Our notation is such that $\rho$ denotes the positive root. The scaling weight for this representation is $c_\alpha=-i\rho$. We can again build a state by acting on the vacuum with this operator
\begin{align}
|-i\rho,\vec{k}\rangle&\equiv|\alpha^H_{N,i\rho}(\vec{k})\rangle\equiv\alpha^H_N(\vec{k})|0\rangle\\
&=\left(\frac{k}{2}\right)^{-i\rho}\sqrt{\frac{\rho\pi}{sinh(\rho\pi)}}\frac{1}{\Gamma(1-i\rho)}|-\vec{k}\rangle.
\end{align}
The tensor product operator $\alpha^H_{N,i\rho_1,i\rho_2}(\vec{k}_1,\vec{k}_2)\equiv\alpha^H_{N,i\rho_1}(\vec{k}_1)\otimes\alpha^H_{N,i\rho_2}(\vec{k}_2)$ gives rise to the following composite state
\begin{align}
|i\rho_1,\vec{k}_1;i\rho_2,\vec{k}_2\rangle&\equiv\alpha^H_{N,i\rho_1}(\vec{k}_1)|0\rangle\otimes \alpha^H_{N,i\rho_2}(\vec{k}_2)|0\rangle\\
&=\frac{\pi}{\sqrt{sinh(\rho_1\pi)\sinh(\rho_2\pi)}}\frac{\sqrt{\rho_1\rho_2}}{\Gamma(1-i\rho_1)\Gamma(1-i\rho_2)}\left(\frac{k_1}{2}\right)^{-i\rho_1}\left(\frac{k_2}{2}\right)^{-i\rho_2}|-\vec{k_1};-\vec{k_2}\rangle.
\end{align}
Noting that $\Gamma(1+i\rho)\Gamma(1-i\rho)=\frac{\pi\rho}{sinh(\pi\rho)}$, this product state is normalizable upto a dirac-delta function
\begin{align}
\langle i\rho_1,\vec{k}_1;i\rho_2,\vec{k}_2|i\rho_1,\vec{k}'_1;i\rho_2,\vec{k}'_2\rangle=\langle -\vec{k}_1;-\vec{k}_2|-\vec{k}'_1;-\vec{k}'_2\rangle=(2\pi)^{2d}\delta^{d}(\vec{k}_1-\vec{k}_1')\delta^{d}(\vec{k}_2-\vec{k}_2').
\end{align}
Thus we have a composite state normalized with respect to the inner product for composition of two principal series representations \eqref{compositeinner} as
\begin{align}
\left(\alpha^H_{N,i\rho_1,i\rho_2},\alpha^H_{N,i\rho_1,i\rho_2}\right)=\frac{1}{\Omega^2}\int\frac{d^dk_1}{(2\pi)^d}\frac{d^dk_2}{(2\pi)^d}\langle i\rho_1,\vec{k}_1;i\rho_2,\vec{k}_2|i\rho_1,\vec{k}'_1;i\rho_2,\vec{k}'_2\rangle=1.
\end{align}
\subsection{Reduction of a composite representation}
To reduce a given representation on $G$ into irreducible ones, one needs a map from $G$ to a product of $\mathcal{V}^{l}$. Such a map happens to exist, as
\begin{align}
Q:~G\to\mathcal{V}^{l_1}\otimes\mathcal{V}^{\tilde{l}_2}.
\end{align}
This map involves the Weyl inversion $w$ and is defined by \cite{Dobrev}
\begin{align}
\left[Q\mathfrak{f}\right](g)=\mathfrak{f}(g,gw).
\end{align} 
The Weyl inversion\footnote{As listed in \cite{Dobrev} chapter 4, Weyl inversion acts on dilatations $a\in A$ as
	\begin{align}
	w^{-1}aw&=waw^{-1}=a^{-1},
	\end{align}
	and on rotations $m\in M$ as
	\begin{align}
	m^w\equiv& wmw^{-1}=w^{-1}mw=\theta m\theta~\text{where}~m^w\in M~\text{for}m\in M,\\
	&D^{\tilde{l}}(m^w)=D^l(m).
	\end{align}
	The subgroups of translations ($\tilde{N}$) and special conformal transformations ($N$) are conjugate to each other under Weyl transformations such that
\begin{align}
w^{-1}n_b w&=wn_b w^{-1}=\tilde{n}_{b'}~\text{where}~b'=(b_1,\dots,d_{d-1},-b_d).
\end{align}
The elements $w$ and the identity together make up the finite group of order two, the Weyl group.}
is based on reflection of the $d^{th}$-axis by
\begin{align}
\text{Weyl inversion:}~~ wx=\frac{\theta x}{x^2},~\text{where}~\theta:\text{reflection of $x_d$}.
\end{align}
From the perspective of reducing representations on $G$ into a product of irreducible representations we are dealing with functions $\mathfrak{f}(g,gw)$. From the perspective of the Kronecker product of irreducible representations we expect these $\mathfrak{f}(g,gw)$ functions to satisfy the covariance condition \eqref{covcond_product} on $\mathfrak{f}(g_1,g_2)$ as if the map $Q$ acted on $\mathfrak{f}(g,g)$. Moreover we mentioned that composite representations are induced by the subgroup $MA$. The function $\mathfrak{f}(g,gw)$ subject to the tensor product covariance condition should schematically work as
\begin{align}
\mathfrak{f}(gma,gmaw)=\left[\dots\right]\mathfrak{f}(g,gw).
\end{align}
To discover the $\left[\dots\right]$ part, making use of the properties of the Weyl inversion, one can rewrite the second argument as follows
\begin{align} gmaw=gmwa^{-1}=gwm^w a^{-1}\end{align}
Then via the covariance condition \eqref{covcond_product}
\begin{align}
\mathfrak{f}(gma,gwm^w a^{-1})&=\left[D^{\chi_1}\left((ma)^{-1}\right)\otimes D^{\chi_2}\left((m^wa^{-1})^{-1}\right)\right]\mathfrak{f}(g,gw)\\
&=|a|^{c_1-c_2}\left[\left(D^{l_1}(m)\right)^{-1}\otimes \left(D^{\tilde{l}_2}(m)\right)^{-1}\right]\mathfrak{f}(g,gw).
\end{align}
Notice that because of the Weyl inversion involved the representation $\chi_2=\{l_2,c_2\}$ works in via $\{\tilde{l}_2,-c_2\}$. The map $Q$ is invertible and it is also an intertwining map.
Defining 
\begin{align}
L(ma)=|a|^{c_2-c_1}\left[D^{l_1}(m)\otimes D^{\tilde{l}_2}(m)\right].\end{align}
we have a space of infinitely differentiable functions 
\begin{align} F:g\to \mathcal{V}^{l_1}\otimes\mathcal{V}^{\tilde{l}_2},\end{align}
with covariance property
\begin{align}
\nonumber\text{covariance condition for $MA$ induced representations:}\\
F(gma)=L(ma)^{-1}F(g).
\end{align}
This function space is denoted as $Q(\mathcal{C}_{\chi_1}\otimes\mathcal{C}_{\chi_2})$ \cite{Dobrev}. A representation that acts on this space is given by the following homomorphism
\begin{align}
\left[\mathcal{T}(g)F\right](g')=F(g^{-1}g).
\end{align}
In the case of purely imaginary $c_1-c_2$, the well defined inner product is \cite{Dobrev} 
\begin{align}
\left(F_1,F_2\right)=\int dn d\tilde{n}\langle F_1(\tilde{n}n),F_2(\tilde{n}n)\rangle,
\end{align}
which is preserved by $Q$,
\begin{align}
\left(QF_1,QF_2\right)=\left(F_1,F_2\right)~~\text{for}~~\chi_1,\chi_2~\text{in principal series}.
\end{align}

In general dimensions, \cite{Dobrevtensor} states that tensor products involving only scalars are reduced in terms of principal series representations only. The absence of discrete series representations in the decomposition is explained in connection to the inequivalence of discrete series representations and their mirror images. In two dimensions, with the $SL(2,R)$ representations the situation is a bit different and even the tensor product of two principal series representations involve discrete series representations, stated by Theorem 4.6 in \cite{Repka}. More recent literature that considers interactions for principal series fields, with the purposes of exploring the  operator product expansion for a dual conformal field theory are  \cite[OPE]. 

\textbf{Acknowledgements:} We thank the participants of Corfu Summer Institute 2021 Workshop on Quantum Features in a de Sitter Universe, for fruitful discussions that have shaped this manuscript and support from  European Union’s Horizon 2020 research and innovation programme
under the Marie Skłodowska-Curie grant agreement No 840709-SymAcc and  European Structural and Investment Fund and the Czech Ministry of Education, Youth and Sports (Project CoGraDS$-CZ.02.1.01/0.0/0.0/15\_003/0000437$) at different stages.

\end{document}